\documentclass[12pt]{article}
\usepackage{amsfonts,amssymb,epsfig,amsmath,graphicx}
\usepackage{float}
\renewcommand{\text}[1]{#1}

\newcommand{\be}{\begin{equation}}
\newcommand{\ee}{\end{equation}}
\newcommand{\ben}{\begin{displaymath}}
\newcommand{\een}{\end{displaymath}}
\newcommand{\bea}{\begin{eqnarray}}
\newcommand{\eea}{\end{eqnarray}}
\newcommand{\bean}{\begin{eqnarray*}}
\newcommand{\eean}{\end{eqnarray*}}
\newcommand{\nn}{\nonumber \\}
\newcommand{\ba}{\begin{array}}
\newcommand{\ea}{\end{array}}
\newcommand{\bi}{\begin{itemize}}
\newcommand{\ei}{\end{itemize}}

\begin{document}
\begin{titlepage}

\vfill
\begin{flushright}
KIAS-P09056
\end{flushright}

\vfill

\begin{center}
   \baselineskip=16pt
   {\Large\bf Compactification driven Hilltop Inflation in Einstein-Yang-Mills.}
   \vskip 2cm
      Eoin \'{O} Colg\'{a}in and Ignacio Zaballa       \vskip .6cm
             \begin{small}
      \textit{Korea Institute for Advanced Study, \\
        Seoul, Korea}
        \end{small}\\*[.6cm]
\end{center}

\vfill
\begin{center}
\textbf{Abstract}\end{center}

\begin{quote}
Starting from Einstein-Yang-Mills in higher dimensions with an instanton on a compact sphere, we dimensionally reduce to find an effective four-dimensional action describing ``hilltop'' inflation. Using recent CMB data, we analyse the parameter space of this model to search for viable set-ups. One unique feature of this class of inflationary models is that the value of the inflaton field, or alternatively, the size of the compact sphere, is stabilised dynamically during the inflationary process.   
\end{quote}

\vfill

\end{titlepage}

\section{Introduction}

With the recent advent of precision cosmology and some imaginative astrophysical projects, we have gained considerable knowledge about the structure of our universe. In particular, we now appreciate that the universe is spatially flat, and the latest 
CMB data from WMAP5 narrows down the curvature perturbation spectral index to $n_s = 0.96 \pm 0.013$  \cite{WMAP}. This data can be plausibly accounted for by an epoch of cosmic inflation in the early universe \cite{inflation}, modelled by a suitable inflaton field $\phi$, whose positive scalar potential $V(\phi)$ powers the nearly exponential expansion.  

The observed temperature fluctuations in the CMB suggest that the shape of the
potential of the inflaton is very flat, at least long before the end of inflation when
the observed fluctuations probe the primordial perturbations generated during 
inflation. 
For a single scalar field driving inflation then, a slow-roll inflationary set-up is
consistent with the current observational data. 
Fast-roll inflation may also be entertained \cite{fastroll}, though one requires at least one
additional field to liberate the inflaton from the task of making a dominant contribution to 
the curvature perturbation \cite{review}.
There are other mechanisms to generate the primordial curvature perturbation
( see for example \cite{others}),
but in this paper we focus on the two mentioned above. 
Other key constraints on these potentials may come from the observation that the present universe is in a state of accelerated expansion \cite{hubble}. As a result, a local minimum of the potential with a small constant vacuum energy $V>0$ corresponding to an asymptotic de Sitter (dS) phase is expected after the end of inflation\footnote{It was argued in \cite{fastroll} that an Anti-de Sitter vacuum cannot be the end result of inflation. For a minimum at $V(\phi)<0$ the field oscillates about the minimum with positive
energy density until the universe reaches its maximum size. After that the universe contracts to eventually collapse into a singularity.}. Finally, the inflationary stage should last about 60 efolds, so that scales corresponding to 
observation have a chance to  leave the horizon during inflation. (See \cite{review}  and references therein for an extensive review of inflation.) 

Although $\phi$ may be a fundamental field, evoking extra-dimensions, other viable candidates for the inflaton emerge, including the location of branes or supersymmetric moduli in a string theory setting.  With extra dimensions, recovering our observable universe, involves typically reducing on a compact space, with the proviso that the extra dimensions are hidden from macroscopic view. Despite the unimaginably large zoo of possible four-dimensional theories one encounters in this process  \cite{Susskind:2003kw}, vibrant developments in observational cosmology, provide a means to whittle them down. On that basis, we have already witnessed many fascinating studies of moduli stabilisation in the string theory literature \cite{moduli}, where CMB data has been used as a powerful probe of the fundamental theory of gravity 
in the early universe.   

This paper builds on the work initiated in \cite{Kihara:2009ea}, where it was demonstrated that compactification in a similar spirit to Cremmer-Scherk ``spontaneous compactification" \cite{Cremmer:1976ir} could arise in a set-up incorporating a non-trivial soliton on the compactified space. In these models, the time-dependent radius of the compactified space may settle down to a finite value, instead of decompactifying to infinity. One bonus feature of this study was the observation that a description of inflation could be incorporated through the compactification process. 

In this paper we will focus on the simpler case where we have a Yang-Mills instanton on $S^4$ and consider compactifications from a d=8 Einstein-Yang-Mills action with gauge-coupling $q_8$ and a positive cosmological constant $\Lambda_8 > 0$\footnote{The Abelian case was presented in \cite{Carroll:2009dn}. They have considerable overlap, though the effective potential varies through the flux terms. }. We will then dimensionally reduce on the $S^4$, before conducting a study of the effective potential and its local maximum/minimum pairing which constitute a ``hilltop" inflation model \cite{hilltop1,hilltop2}. Finally, we explore the stability of the de Sitter vacuum to general metric fluctuations to see if tachyonic instabilities further constrain the parameter space. 

\section{Set-up}

We start from considerations of the following action in d=8 space-time,
\bea
S_8 &=& \int d^{8}x \sqrt{-g_8} \left( \frac{\mathcal{R}}{16 \pi G_8} - \frac{\Lambda_8}{2} - \frac{1}{4} ( F^{a}_{mn} F^{a \; mn}) \right),
\eea
where $\Lambda_8$ and $G_8$ are the d=8 cosmological constant and Newton's constant respectively, while $a=1,2,3$ are the indices on the $SU(2)$-valued instanton field strength $F^a$ on the $S^{4}$. $F$ may in turn be expressed in terms of the gauge potential as
\be
\label{fstr}
F = dA + {{q_8}} A \wedge A,
\ee
where $q_8$ is the gauge coupling constant. In terms of projective coordinates $y^{m}, m=1,...,4$, we may express the metric on the $S^4$ as
\be
ds^{2}(S^4) = \frac{\delta_{mn} dy^{m} d y^{n}}{(1+ |y|^2/4)^2},
\ee
with $|y|^2 =  y_{m} y^{m}$.

We may now introduce a gauge configuration which satisfies the Bogomol'nyi duality relation $*_{4} F = \pm F$. We consider a static gauge configuration on the $S^4$ with gauge potential 
\be
\label{pot}
A =  \frac{1}{4 q_8} \frac{|y|^2}{(1 + |y|^2/4)} U^{\dag} d U,
\ee
where we define $U$ to be
\be
U = \frac{1}{|y|} Y, \quad Y = y^4 {\mathbf{1}}_2 + y^1 \tau_1 + y^2 \tau_2 + y^3 \tau_3.
\ee
The matrices $\tau_{a}$ are related to the usual Pauli-matrices, $\tau_{a} = -i \sigma_{a}$, and satisfy the multiplication relation $\tau_{a} \tau_{b} = -\delta_{ab} \mathbf{1}_{2} + \epsilon_{abc} \tau_{c}$. The field strength corresponding to  (\ref{pot}) may be written
\be
F =  \frac{1}{4 q_8}  \frac{1 }{ (1 + |y|^2/4 )^2}  d  Y^{\dag} \wedge  d Y,
\ee
or, alternatively, in terms of components $ F = F^{a} \tau_{a}$:
\bea
F^a &=& -\frac{1}{4 q_8} \frac{ \epsilon^{a b c} dy^{b} \wedge d y^{c}+2 dy^{a} \wedge dy^{4}}{(1+ |y|^2/4)^2}. 
\eea
One may readily check that the gauge configuration is self-dual: $ F = *_{4} F$. The equations of motion are then ensured via the Bianchi identity, which is satisfied by construction.

\subsection{The d=4 effective potential}


Although we may consider the equations of motion in d=8, a better view of the dynamics of the system may be gleaned by reducing on the $S^4$ and looking at the effective potential in d=4 in Einstein frame. For simplicity, we just exhibit the low-energy effective action
for the lowest Kaluza-Klein mode of $\phi$, with higher modes being discussed in section 4. To dimensionally reduce, we introduce the following time-dependent ansatz for the metric
\bea
\label{met_ansatz}
ds^{2} =  e^{-4 \phi(t)} g^{(4)}_{\mu \nu} dx^{\mu} dx^{\nu} + e^{2 \phi(t)} ds^{2} (S^4),
\eea
where the $g_{\mu \nu}^{(4)}$ denotes the usual metric of a Friedmann-Robertson-Walker (FRW) spacetime,
\be
g^{(4)}_{\mu \nu} dx^{\mu} dx^{\nu} = -dt^2 + a^2(t)[dr^2 + r^2 d \Omega^2]. 
\ee 
Denoting the radius of $S^4$ by $L_0$, the reduced action then becomes  
\be
S_4 = \int d^{4} x \sqrt{-g_4} \left( \frac{1}{16 \pi G_4} [ R_4 + 4 \nabla_{\mu} \nabla^{\mu} \phi - 12 \partial_{\mu} \phi \partial^{\mu} \phi ] -V_4 \right),
\ee
where 
\be
V_{4} = \frac{\Lambda_4}{2} e^{-4 \phi} + \frac{3}{4 q_4^2 L_0^4} e^{-8 \phi} - \frac{3}{4 \pi G_4 L_0^2} e^{-6 \phi},  
\ee
and we have redefined new d=4 parameters in terms of the volume of $S^4$, $V_{S^4} = {\pi^2 L_0^4}/{2}$:
\be
 \quad G_{4} = \frac{ G_{8}}{ V_{S^4}} , \quad \Lambda_4 = V_{S^4} \Lambda_8, \quad q_{4} = \frac{q_8 }{\sqrt{V_{S^4}}}. 
\ee

Note above that the second term in the action is a total derivative term, and may be integrated out\footnote{This may be easily seen by bearing in mind that the covariant derivative the scalar density $\sqrt{-g_4}$ may be given as 
$
\nabla_{\mu} \sqrt{-g_4} = \partial_{\mu} \sqrt{-g_4} - \Gamma^{\nu}_{~\nu \mu} \sqrt{-g_4} = 0. $ }. 
Finally, bringing the action to the canonical form, requires the rescaling $\phi \rightarrow \tfrac{1}{2 \sqrt{3} M_p} \phi$, 
\be
S= \int d^{4} x \sqrt{-g_4}  [\frac{1}{16 \pi G_4} R - \frac{1}{2} \partial_{\mu} \phi\partial^{\mu} \phi -{V}_4],
\ee
where now 
\be
{V}_{4} = \frac{\Lambda_4}{2} e^{-2 {\phi}/(\sqrt{3}M_p)} + \frac{3}{4 q_4^2 L_0^4} e^{-4 \phi/(\sqrt{3}M_p)} - \frac{3}{4 \pi G_4 L_0^2} e^{-3 \phi/(\sqrt{3}M_p)}. 
\ee
As is customary in the literature, we have used the reduced Planck mass $M_p = (8 \pi G_4)^{-1/2}$ to absorb factors appearing in the Einstein equations.
In preparation for later numerical work, we also use this opportunity to introduce the dimensionless paramaters 
\be
\label{dimensionless}
b \equiv \frac{4 \pi G_4}{q_4^2 L_0^2}, \quad c \equiv 4 \pi G_4 \Lambda_4 L_0^2,
\ee
so that the potential becomes 
\be
\label{epot}
V_4 = \frac{M_{p}^2}{L_0^2} \left[ c e^{-2 \phi/(\sqrt{3} M_p)} + \frac{3 b}{2} e^{-4 {\phi}/(\sqrt{3} M_p)} - 6 e^{-3 {\phi}/(\sqrt{3} M_p)} \right]. 
\ee 
From now on we will refer to $V_4$ simply as $V$. 

We now proceed to examine this effective potential. ${\phi} \rightarrow \infty$ corresponds to the $S^4$ decompactifying, so we ignore this possibility. Also, in the opposite direction, as $b>0$, $V$ goes to infinity for small ${\phi}$. In order to stabilise ${\phi}$ and ensure that it doesn't run away, we require $V$ to have two extrema. This means that the solution to $V' = 0$,
\be
\phi = -\sqrt{3}M_p\, {\rm ln}\left[\frac{3}{2b}
\left(1 \pm \sqrt{1-\tfrac{4bc}{27}}\right)\right]
\,,
\ee
should correspond to two roots, which in turn dictates the upper bound on the 
product of $b$ and $c$ is $bc < \tfrac{27}{4}$.
These roots now correspond to a minimum $\phi_{min}$ and a maximum $\phi_{max}$ respectively, and offer the ideal place to look for an inflationary model, in which inflation takes place at the top of the 
potential: hilltop inflation \cite{hilltop1,hilltop2}. For later use, we also here record the distance between the two roots
\be 
\label{phidiff}
\phi_{max} - \phi_{min} = - 2 \sqrt{3} M_p \ln \left[ \sqrt{\tfrac{27}{4 bc}} \left(1- \sqrt{1-\tfrac{4bc}{27}} \right)\right],  
\ee
and also note explicitly the combination $bc$ in terms of the original $d=8$ parameters of the model
\be bc = \frac{(4 \pi G_8)^2 \Lambda_8}{q_8^2}. \ee

\begin{figure}[H]
\label{potentials}
 \begin{center}
  \includegraphics[scale=0.9]{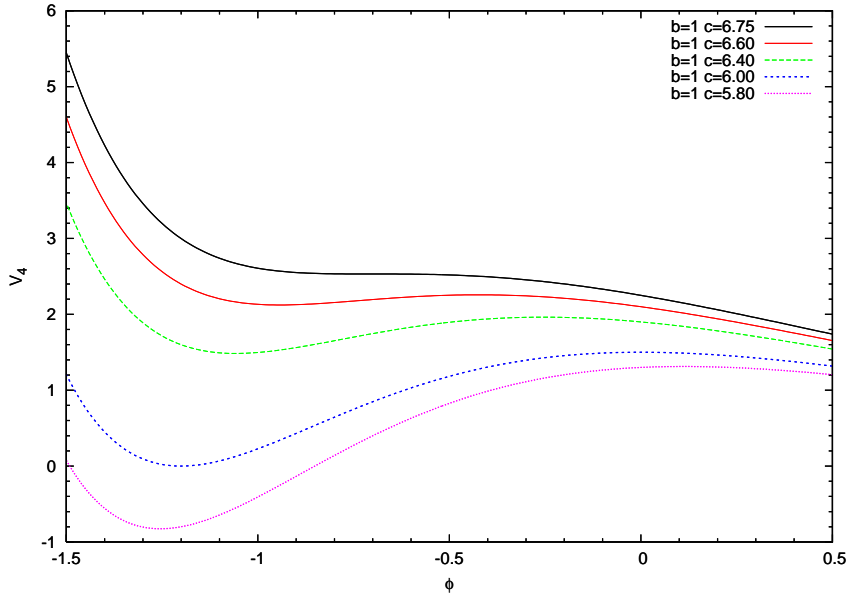}
 \end{center}
\vspace{-5mm}
\caption{\scriptsize{The figure shows the effective 4d potential $V$ for $b=1$ and 
different values of  the product $bc$. The  effect of increasing $bc$ corresponds to 
the raising the de Sitter vacuum. }}
\end{figure}

Another constraint on the ($b,c$) parameter space comes from looking at the roots of the potential $V = 0$:
\be
\phi = -\sqrt{3}M_p\, {\rm ln}\left[\frac{2}{b}
\left(1 \pm \sqrt{1-\tfrac{bc}{6}}\right)\right].
\ee
Avoiding a potential with a negative minimum, one corresponding to an Anti-de Sitter vacuum,  necessitates $bc \geq 6$, making the final range for compactification to a Minkowski/de-Sitter vacuum to be 
\be
\label{window}
6 \leq bc < \tfrac{27}{4}. 
\ee
Within this range, the effect of varying $bc$ may be noted from Fig. (\ref{potentials}), where the vacuum raises from Anti-de Sitter ($bc < 6$) to Minkowski ($bc=6$) and de-Sitter ($bc>6$).

Within the permitted range for the parameters (\ref{window}), we note from Fig. (\ref{potentials}) that there are three channels for the dynamics of the scalar field
$\phi$ depending on its initial conditions.
Denoting $\phi_{min}$ and $\phi_{max}$ to be the value of the field at the local minimum and maximum respectively, the three channels correspond to $\phi < \phi_{min}$, $\phi_{min} < \phi < \phi_{max}$, and $\phi > \phi_{max}$, where we have omitted the classically stable extrema from the quoted ranges. Our focus in this paper will be the second channel, though some amount of inflation may occur in the other two.  

We ignore the third possibility, as from the reduction ansatz, the run-away direction for $\phi$ corresponds to the extra dimensions decompactifying. The first channel is also constrained and there is an lower bound on the value of $\phi$ in order to avoid a scenario where the
field $\phi$ reaches the minimum with enough kinetic energy to surmount the local maximum and also escape to infinity. We show solutions for the evolution of $\phi$ via this channel later in the text. 

Therefore, in order to safely stabilise the compactified extra dimensions, we focus on the case where 
the inflaton rolls down to the minimum (stable vacuum) of the potential from somewhere near the maximum (unstable vacuum) giving rise to inflation. In contrast to work initiated in \cite{Kihara:2009ea} where compactification in these models was studied at the level of the equations of motion in higher dimensions, the reduced action and effective potential, offer a more intuitive picture, and also allow an estimation of the primordial perturbations produced during inflation to compare with CMB observational data. 


\section{Inflation}
\setcounter{equation}{0}

In this section, we analyse the effective potential in (\ref{epot}) in terms of 
the slow-roll parameters $\epsilon$ and $\eta$, to determine how the values of the paramters $b$ and $c$ affect the resulting nature of the inflation. 
Our purpose here is to explore the effective action as a viable model of inflation, identify any potential shortcomings that may arise, and address those concerns in future work. 
Before proceeding to the analysis, we summarise the basics of slow-roll inflation 
and highlight the current observational bounds that are the ultimate judges of any 
inflationary model. 

For a homogeneous inflaton field, $\nabla \phi = 0$, the equation of motion for the 
field $\phi$ is
\be \label{eoms}
\ddot{\phi} + 3 H \dot{\phi} + V'(\phi)=0
\,,
\ee
where $H^2=\rho/3 M_p^2$ and $\rho$ is the energy density of the 
inflaton field.
In the slow-roll approximation, $\dot{{\phi}}^2 \ll V({\phi})$, the
equation of motion for the field is approximately given by
$
3 H \dot{{\phi}} \simeq -V'(\phi)
$
with $H^{2} \simeq V({\phi})/3 M_p^2$.
A necessary condition for the slow-roll approximation to be valid is
that the slow-roll parameters $\epsilon$ and $\eta$, defined by
\be
\epsilon = \frac{M_p^2}{2}  \left( \frac{V'}{V} \right)^2, \quad \eta = M_p^2 \frac{V''}{V}, 
\ee
are small, $\epsilon\ll1$ and $\vert\eta\vert\ll1$ \cite{Liddle:1992wi,Liddle:1993fq}.
In our hilltop model if the slow-roll conditions are satisfied, then 
$\epsilon\ll\vert\eta\vert$. 
The end of inflation coincides with  $\epsilon = 1$, with $\epsilon < 1$ being a  
necessary condition for inflation to occur. As a result, by relaxing the slow-roll conditions, fast-roll inflation \cite{fastroll} is also permitted when $|\eta| \sim O(1)$. We investigate both of these possibilities later.  

The vacuum fluctuations of the inflaton field generate, at horizon
exit, the curvature perturbation ${\cal R}$.  Its amplitude and 
scale dependence are typically described in terms of the power spectrum
and the spectral index. 
At the time of horizon exit during inflation, these are given respectively by
\be
\label{Pr}
\mathcal{P}_{\mathcal{R}} = \frac{1}{24 \pi^2 M_p^4} 
\left(\frac{V}{\epsilon}\right)
\,,
\ee
and
\be
n_s-1 =  2 \eta -6 \epsilon\simeq 2 \eta_0
\,.
\ee
The approximate expression for the spectral index is the result
of the slow-roll approximation for inflation in the proximity of the maximum 
of the potential.
The subscript `$0$' denotes the evaluation near the maximum of the 
potential. 
In addition to density perturbations, tensor perturbations or gravitational
waves are produced during the inflationary epoch. The scalar to tensor 
fraction is given by $r=16\epsilon$.  

According to five-year WMAP data \cite{WMAP}, the observed amplitude
of the curvature perturbation  and spectral index are
\bea
\label{obs_spec}
n_s &=&  0.960\pm0.013, \\
\label{obs_index}
\mathcal{P}_{\mathcal{R}}&=& 2.5 \times 10^{-9}. 
\eea
Neglecting any fine tuned cancellations between $\epsilon$ and $\eta$, and the absence of secondary fields, 
the observed value of the spectral index can only be realised in a scenario
where both of these parameters are small. 
Additionally, the absence of tensor perturbations in the CMB
places an upper bound on $r$, $r<0.2$.
This upper bound constrains the scale of inflation to be
\be\label{Vrbound}
V^{1/4} < 2 \times 10^{16} \, \mbox{GeV}
\,.
\ee 

If we adopt the maximum inflation scale and consider instant reheating,
we require about $N=60$ efolds of inflation for the scales corresponding to 
observation to leave the horizon during inflation. 
Reducing the scale of inflation and delaying reheating after the end
of inflation reduces the number of efolds required \cite{Alabidi:2005qi}.

As we have a hilltop model, a great deal may be understood from expanding the potential around the local maximum. 
Expanding the potential to second order about the local maximum, one may write 
\be\label{quadratic_exp}
V(\phi) \simeq V_{0} - \frac{m^2}{2} (\phi-\phi_{max})^2+ \cdots, 
\ee
where the remaining terms are deemed to be negligible.  
Here $V_{0}$ denotes the potential energy at the maximum, and 
the vacuum expectation value of $\phi$ is of the order of $M_p$. 
In the context of our model, the above constants depend on $b$ 
and $c$:
\bea
\label{quad_constants}
V_{0} &=&  \frac{M_{p}^2}{L_0^2} \frac{(-9+\sqrt{81-12 bc})^2}{144 b^3} (2 bc-9 + \sqrt{81 -12 bc}), \nn 
m^2 &=& \frac{1}{L_0^2} \frac{(-9+\sqrt{81-12 bc})^2}{108 b^3} (4 bc - 27 +3 \sqrt{81-12 bc }).  
\eea

To this degree of approximation, the slow-roll parameters become 
\bea \label{approx}
\epsilon &=& \frac{(\phi - \phi_{max})^2}{2 M_p^2} \eta_0^2, \nn
\eta_0 &=& \frac{4}{3} \frac{(4 bc - 27 +3 \sqrt{81-12 bc })}
{(2 bc-9 + \sqrt{81 -12 bc})}. 
\eea

At the top of the potential $|\eta_0|$ varies as $bc$ moves in its permitted 
range $6 \leq bc < 27/4$ as
\be
\label{eta0}
\frac{4}{3} \geq |\eta_{0}| = \frac{4}{3}\frac{(4 bc - 27 +3 \sqrt{81-12 bc })}
{(2 bc-9 + \sqrt{81 -12 bc})} > 0
\,,  
\ee
with $|\eta_0| \simeq 4/3$ correspond to small $bc$, with increasing values of the product $bc$ leading to smaller and smaller values of $\eta$.  As may be seen above, $|\eta_0|=1$ corresponds to $bc = 162/25 = 6.48$, thus delineating where the slow-roll approximation ends and a fast-roll description becomes valid. 

One more special value of $bc$ requires mention. As can be seen from (\ref{phidiff}), as $bc$ increases, the difference between the extrema decrease until one reaches $bc=27/4$ where they coalesce. For large enough $bc$, as may be appreciated from (\ref{approx}), $\epsilon$ may never reach unity to end inflation. To the quadratic order above, once $bc$ exceeds $6.1$ inflation fails to end. 

The above values $bc \simeq 6.1$ an $bc \simeq 6.48$ we have produced by expanding the potential to  quadratic order. However, alternatively one may work directly with the potential (\ref{epot}) and drop the approximation. Taking into account that $\epsilon, \eta$ are curves with maximum and a minimum resepectively, it is possible to repeat the simplified analysis. In general, inflation ends provided $bc \lesssim 6.21$ and $bc \simeq 6.49$ separates slow-roll from fast-roll inflation. 

In the next two subsections we take a closer look at these regions. In each case we
may determine the relevant quantities analytically when it is possible for a comparison
with numerical results. 

\subsection{Slow-roll Inflation} 

In this subsection, we consider slow-roll solutions for the effective
potential in (\ref{epot}).
In this class of models $\vert\eta\vert\gg\epsilon$ along the 60 or
so efolds of inflation that occur close to the maximum of the 
potential. 

The slow-roll parameter $\eta$ is less than the unity
at the maximum of the potential for values of the parameters $b$ and 
$c$ such that  $bc\gtrsim6.49$
As we increase the value of $bc$ for $b$ fixed,
the potential energy between the stable and unstable 
vacua  and the distance between the two extrema decrease.
As $bc$ approaches the upper limit $bc=27/4$, $\eta$
becomes negligibly small and tends to zero.
It is close to this value of $bc$ that we obtain slow-roll
solutions.

Numerical solutions for the equation of motion of $\phi$, (\ref{eoms}), may be found by introducing the following
dimensionless variables 
\be
\bar{\phi} = \frac{\phi}{M_p}, \quad \bar{t} = \frac{t}{L_0}
\quad V = \frac{L_0^2}{M_{p}^2} {V}. 
\ee

\begin{figure}
\begin{center}
\includegraphics[angle=0,width=1.03\textwidth]{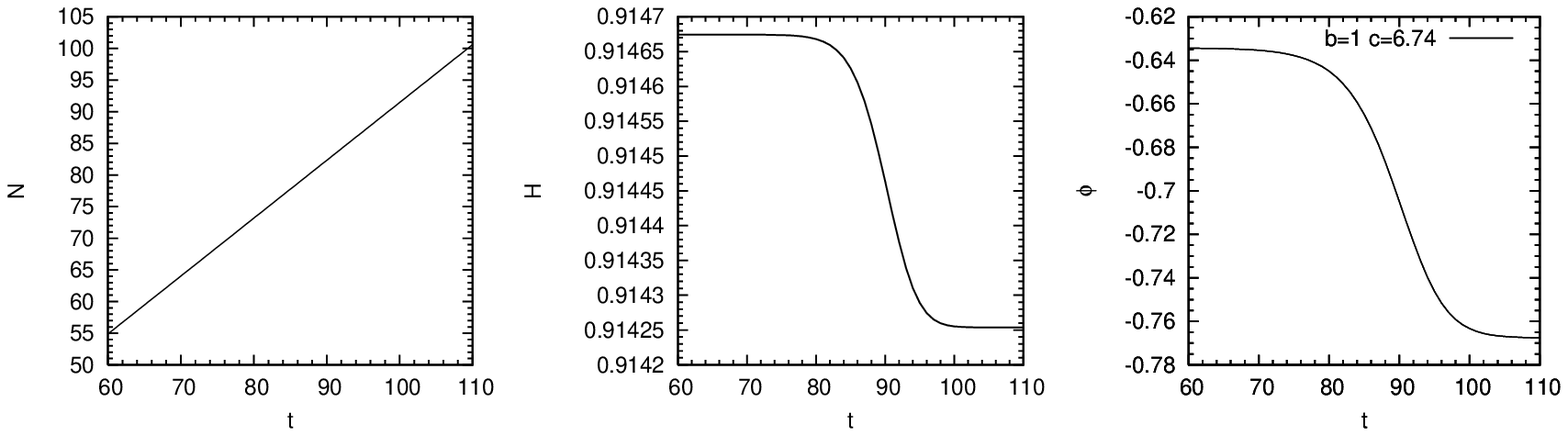}
\end{center}
\caption{\label{Fig:2}
\scriptsize In the figures above we show the number of efolds $N$, the 
Hubble parameter $H$, and the evolution of the field $\phi$ as a function
of time $t$ in $L_0$ units for $b=1$ and $c=6.74$. 
$H$ is given in $L_0^{-1}$ units and $\phi$ in $M_p$
units. In this case we obtain fast-roll at the top of the potential with 
$\vert\eta_0\vert\ll1$.We set the initial conditions close enough to the maximum
of the potential to ensure that we get about $55$ efolds of inflation about
the top of the potential.}
\end{figure}

In Fig.(\ref{Fig:2}) we show the number of efolds $N$, the
Hubble parameter $H$ and the evolution of the field $\phi$
for the case $bc=6.74$. The field starts rolling down in the 
proximity of the maximum of the potential with negligible
velocity to ensure that the slow-roll conditions are
satisfied.
The Hubble parameter remains practically constant as the field 
$\phi$ slowly rolls close to the maximum of the potential. 
The number of efolds of inflation at the top of the potential
is about $N=70$ in this particular case shown in Fig.(\ref{Fig:2}).
The field eventually falls into the stable vacuum, but without 
sufficient kinetic energy to enter a stage of coherent oscillations 
and reheating.
The change in $H$ as the field evolves from the top of the 
potential to the minimum is very small, $\dot H/H^2\ll1$, 
and therefore the slow-roll conditions are never violated throughout
the evolution of $\phi$.
Once the field settles down at the minimum of the potential
we obtain de Sitter accelerated expansion with $V$ constant.

The dynamics of the field for the case $bc=6.74$ described 
above, is similar to all the models of our effective potential
with values of the parameters $b$ and $c$ such that 
$bc\gtrsim6.49$.
Since $\epsilon$ never reaches unity as the field evolves to
its stable vacuum inflation continues at a different scale, 
determined by the actual value of $bc$, after the field reaches the 
minimum of the potential. 
The introduction of a secondary waterfall field with a large mass would 
ensure that reheating occurs very rapidly while the slow-roll conditions 
for the dynamics of the field $\phi$ hold at the end of inflation, like in hybrid
inflationary models
\cite{Linde:1993cn}.
Ideally, the extra waterfall field would come from dimensional reduction
of the d=8 Einstein-Yang-Mills theory. The analysis of this possibility for 
our extra-dimensional theory is however beyond the scope of this 
preliminary paper. We hope to address it in follow-up work. 

Despite this model of slow-roll inflation from our extra dimensional
theory not being in itself a complete realistic model, it is interesting to
investigate the relation between the size of the extra dimensions 
and the observational bounds from the CMB. To do so we assume 
that inflation occurs at the top of the potential, and that it ends 
about 60 efolds after the cosmological scales corresponding to 
the observed CMB leave the horizon during inflation.   

For inflation at the top of the potential, $H$ is practically constant.
The equation of motion for $\phi$ is easily solved expanding the
potential to quadratic order and taking $H$ constant.
The evolution of $\phi$ is given by \cite{fastroll}
\be\label{phislow}
\Delta\phi_N = \Delta\phi_f \, e^{-FN}
\,,
\ee
where $\Delta\phi_N\equiv\phi(N)-\phi_{max}$, 
and
\be
F \equiv \frac{3}{2}\left(  \sqrt{1 + \tfrac{4}{3} | \eta_0|} - 1\right)
\,. 
\ee
$\phi_f$ here denotes where the approximation 
$\eta(t)\simeq\eta_0$ breaks down and the quadratic
expansion is not longer valid.
The analytical approximation agrees very well with our numerical
estimations. 
For $bc=6.74$, $\vert\eta_0\vert=0.276$.
From Fig. (\ref{Fig:2}) we get $\vert\Delta\phi_f\vert\simeq0.011M_p$
and $N=73$ for $t=80L_0$ , before $H$ and $\eta$ change substantially. 
For this values we get $\vert\Delta\phi_N\vert\simeq10^{-10}M_p$, 
which is precisely the value we used in our 
numerical computation as the initial conditions for $\phi(t)$.

In these models, we get a small $\vert\eta_0\vert$ 
compatible with the observed spectral index for values of $bc$
very close to the limit $bc=27/4$. 
Taking a small increment $\delta$
such that $bc=27/4-\delta$, the observational constraint on $n_s$
requires values of $\delta$ in the range 
$\delta=2\times10^{-5}-8\times10^{-5}$.
Then the variation of the field $\phi$ is approximately given by
$
\Delta\phi_N = \Delta\phi_f \, e^{-\vert\eta_0\vert N}
$\,,
where $\vert\eta_0\vert\simeq M_p^2m^2/3V_0$
with $F\simeq\vert\eta_0\vert\ll1$.
On the other hand, using the observed value of $\mathcal{P}_{\cal R}$, 
the scale of inflation is given by
\be
V_0^{1/4}=2.3\times10^{-2}M_p
\left(\frac{\vert\Delta\phi_f\vert}{M_p}\right)^{1/2}
e^{-\vert\eta_0\vert N/2}
\,.
\ee
In any event, the quantity $\vert\Delta\phi_f\vert$ can not be larger than the 
distance between the maximum and minimum for inflation occurring
at the maximum of the potential. An approximate estimate then
for $V_0^{1/4}$, given the allowed range of values of $\delta$ which
gives about $\vert\Delta\phi_f\vert<0.01M_p$, is 
\be
\label{v0bd}
V_0^{1/4}<1.8\times10^{-4}M_p=2\times10^{15}\, \mbox{GeV}
\,,
\ee
where we have used $N=60$ in (\ref{phislow}) for an estimate
of $\Delta\phi_N$. The scale of inflation in this case 
is below the upper bound in (\ref{Vrbound})
from the absence of tensor modes in the CMB, as it was shown
in  \cite{hilltop1} for $\vert\Delta\phi_f\vert\lesssim M_p$. 


\subsection{Fast-roll inflation}
\setcounter{equation}{0}


Apart form the slow-roll solutions for the 4-dimensional effective action,
there is a broad range of values of $bc$ for which we get inflationary
solutions with $\eta\simeq O(1)$. This regime corresponds to fast-roll
inflation \cite{fastroll, Linde:1993cn}.  
In fast-roll inflation values of $\eta\simeq O(1)$ are too large to
match the observed spectral index in the CMB. Then to have a viable
model of inflation, it is necessary that some additional field  
releases the inflaton field from generating the curvature perturbation8
(see e.g \cite{review}). The inflaton may still contribute to the 
total amplitude of the curvature perturbation, but the spectral index
would not be related to the perturbations generated by the inflaton.
For $bc\lesssim6.21$, in addition
to getting values $\eta_0\simeq O(1)$, inflation ends when 
$\Delta\phi_e\equiv(\phi_e-\phi_{max})\simeq M_p$ corresponding to
$\epsilon\simeq O(1)$. Below we focus on this particular case.

\begin{figure}
\begin{center}
\includegraphics[angle=0,width=1.03\textwidth]{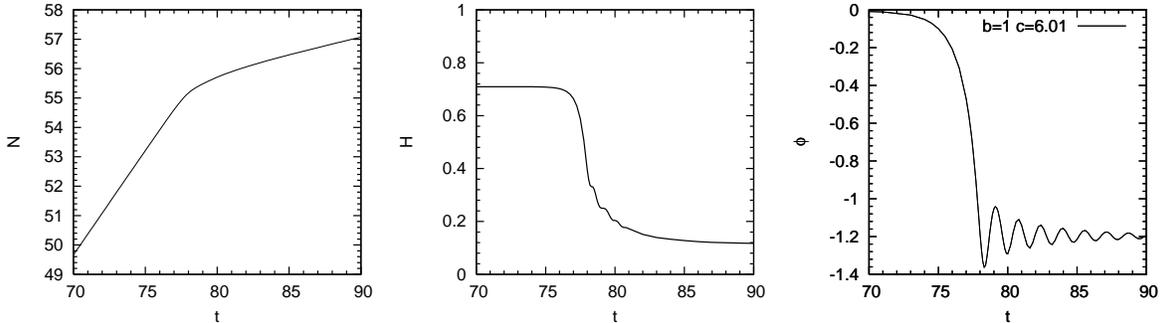}
\end{center}
\caption{\label{Fig:3}
\scriptsize In the figures above we show the number of efolds $N$, the 
Hubble parameter $H$, and the evolution of the field $\phi$ as a function
of time $t$ in $L_0$ units for $b=1$ and $c=6.74$. 
$H$ is given in $L_0^{-1}$ units and $\phi$ in $M_p$
units. In this case we obtain fast-roll at the top of the potential with 
$\vert\eta_0\vert\simeq O(1)$.  We set the initial conditions close enough to the maximum
of the potential to ensure that we get about $70$ efolds of inflation about
the top of the potential.}
\end{figure}

Again, we can reproduce the dynamics of the field $\phi$ numerically. 
In Fig.(\ref{Fig:3}) we show numerical solutions of the equations of
motion for $bc=6.01$.
As in the case in  the previous section, the Hubble parameter remains
practically constant as the field $\phi$ rolls down in the proximity of the
maximum of the potential. The field eventually falls into the stable vacuum
as well, but then it enters a regime of coherent oscillations.
The condition for coherent oscillation for the field at the minimum 
of the potential is $V''\gg H^2$ or $\eta\gg1$, which is satisfied in this
case. Then one expects that during the coherent oscillations, the field $\phi$
decays due to quantum particle creation of other fields that might 
couple to $\phi$. The damping of the oscillations leads to the reheating
of the universe \cite{Kolb:1990vq}.
We find about $N=55$ efolds of inflation on the top of the potential
before the field enters the coherent oscillations period in this case.
Then it behaves as matter domination, as it is shown in  Fig. (\ref{Fig:3}).

The variation of $V$, or alternatively $H$, from the top of the potential
to the minimum increases as $bc$ approaches to the limit value $6$.
For $bc=6$ we have $V=0$ at the minimum, which corresponds to
the Minkowski vacuum. 
Expanding the potential at the minimum about the value $bc=6$,
we get
\be
\frac{L_0^2}{M_p^2}V_{min}\simeq4\delta+O(\delta^2)
\,,
\ee
where $\delta>0$ and it represents a small quantity. 
A stable vacuum energy similar to $\Lambda\sim10^{-120}M_p^4$
to reproduce the observed dark energy requires tuning the value
of $\delta$ very close to $0$.

The analytical solution for fast-roll is similar to the ones we 
derived in (\ref{phislow}). That is
\be
\Delta\phi_N = \Delta\phi_e e^{-F N},
\ee
where $\Delta\phi_N\equiv\phi(N)-\phi_{max}$ and  in this case
$\phi_e$ denotes the end of inflation.
One can in principle take the classical value of $\Delta\phi_N$ as close as necessary to 
the value of the field at the maximum to get an arbitrarily large number of efolds 
of inflation\footnote{However, as argued in \cite{fastroll}, if the classical displacement of the field is subject to quantum fluctuations, there is a lower bound for
$\Delta\phi_N$ that amounts to $\Delta\phi_N\sim m$ for 
$\vert\eta\vert\simeq O(1)$}.
In the numerical estimation in Fig. (\ref{Fig:3}) the initial value of
the field is $\Delta\phi_N=10^{-24}$.  
The end of inflation is determined by $\epsilon = 1$, which gives
$\Delta\phi_e\simeq\sqrt{2} M_p$. For $bc=6.01$ $\eta_0=1.34$.
Then the number of efolds of fast-roll inflation is 
\be
N\simeq\frac{1}{F} \ln\left(\frac{\Delta\phi_N}{ \Delta \phi_e}\right)
\simeq55
\,, 
\ee
which is approximately the number of efolds we have got in the
numerical estimation shown in Fig~.(\ref{Fig:3}). 
In this case the analytical approximation is quite accurate 
reproducing our numerical results.

\subsection{Some remarks}

It would be nice if the size of the extra dimensions were constrained sufficiently by the CMB data so that we could make a prediction. They are insufficient, but the size of the extra dimensions, $L_{ex}$, can be determined from (\ref{met_ansatz}) and  $\phi_{min}$ in terms of a single parameter: 
\be 
L_{ex} \simeq \sqrt{b} L_0 \simeq \frac{1}{M_p q_4}. 
\ee 
Some added assumptions are required to go further.

\begin{figure}
\begin{center}
\includegraphics[angle=0,width=1\textwidth]{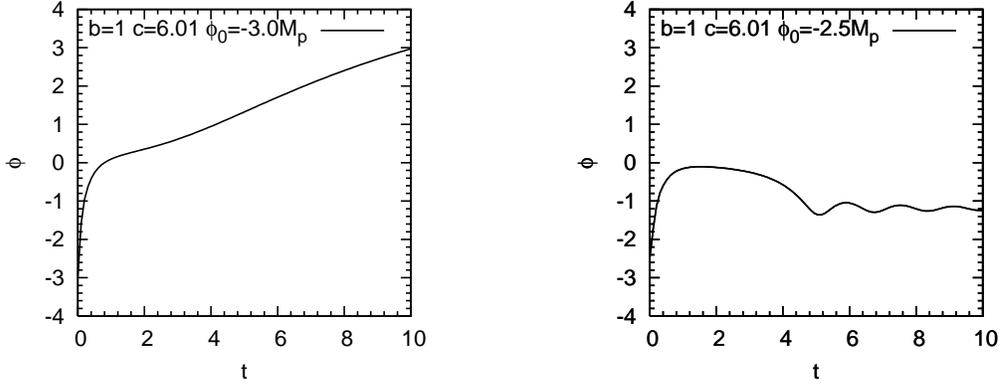}
\end{center}
\caption{\label{Fig:4}
\scriptsize In these figures we the existence of a critical value of $\phi$ in the left branch of inflation. In the first figure, the inflaton rolls over the ``bump`` presented by the local maximum and escapes, whereas, in the second figure, we see compactification at work. As in the previous figures $\phi$ is given in $M_p$ units whereas $t$ in units of $L_0$.}
\end{figure}

 As explained in the introduction, we have opted to focus on inflation arising from the inflaton traveling between the extrema i.e. $\phi_{min} < \phi < \phi_{max}$. However, referring to Fig. (\ref{potentials}), and neglecting $\phi> \phi_{max} $ where the size of the $S^4$ is not stabilised, one remaining candidate for inflation with compactification is an inflaton field that starts to the left of the minimum, $\phi < \phi_{min}$. From Fig.~(\ref{Fig:4}), there is a lower bound on the value of $\phi$ so that the sphere does not decompactify.  Beyond this critical value, the field $\phi$ is not prevented from escaping to infinity. However, as can be seen from Fig. (\ref{Fig:4}),  a significant amount of inflation in this channel may only occur
once the field $\phi$ stabilises at the minimum.

One additional comment needs to be made to conclude this section. As our inflaton starts off close to the top of a hill, $\epsilon$ is very small, and eternal inflation \cite{eternal} is a possibility in all these models. As explained in a recent review \cite{eternal_review}, as the inflaton field $\phi$ rolls down the hill, the change in its classical value $\Delta \phi_{cl}$ in time $\Delta t$,  is subject to additional quantum fluctuations i.e. $\Delta {\phi} = \Delta \phi_{cl} + \Delta \phi_{q}$. In the ordinary, non-eternal regime, quantum fluctuations are just small corrections to the classical story and may be ignored. However, if the classical $\dot{\phi}$ becomes very small - smaller than $H^2$ - the classical inflaton change drops below the typical size of quantum fluctuations. As a result, there is finite probability that the inflaton field will move up the hill, meaning that over the interval $\Delta t$, regions with these larger values for $\phi$ will expand more than the average. As this process is repeated for successive time intervals $\Delta t$, one ends up with an infinite number of universes, providing an echo of the multiverse \cite{Susskind:2003kw} appearing in string theory. 

\section{Vacuum stability}
\setcounter{equation}{0}

In general there is no guarantee that the vacuum at the end of inflation is stable against tachyons. Here, we address the constraints that arise from demanding that the final vacuum is stable against metric fluctuations to linear order in the d=8 Einstein equations. As the compactified vacuum corresponds to $\dot{\phi} = 0$, the overall spacetime becomes a direct product of a FRW spacetime and a four-sphere and there is some simplification.  
 
To begin with, the d=8 equations of motion may be re-written 
\be
\label{neweom}
\frac{R_{MN}}{16 \pi G} =   F^{a}_{MP} F_{a \; N}^{~P} - \frac{g_{MN}}{12} \left( \frac{1}{2} F^{a}_{MN} F^{a \; MN} - \Lambda\right),
\ee
where $M,N = 1,...,8$. 

Now we consider fluctuations in the metric $\delta g_{MN}$, but will assume for simplicity that the gauge field is not varied $\delta A = 0$. We will follow the treatment which parallels work in \cite{DeWolfe:2001nz,Bousso:2002fi},  using Greek indices for the non-compact and Roman for the compact space. The basic idea is to vary the Einstein equations above to the linear level in $\delta g_{MN}$. These fluctuations may be expanded in terms of the spherical harmonics on $S^4$, $Y^{I}$. The idea then is to check that there is no tachyonic instability by ensuring that the mass squared term in the resulting equations 
\be
(\Box_{FRW} - m^2) Y^{I} = 0, 
\ee
is non-negative, and thus of the correct sign. If $m^2 <0$, this would herald the onset of a tachyonic instability. 

Proceeding, we firstly consider the following linearised metric fluctuations
\bea
&&\delta g_{\mu \nu} = H_{\mu \nu} - \frac{1}{2} g_{\mu \nu} h^{c}_{c}, \quad H_{\mu \nu} = H_{(\mu \nu)} + \frac{1}{4} g_{\mu \nu} H^{\rho}_{\rho}, \nn && \delta g_{\mu a} = h_{\mu a}, \quad
\delta g_{a b} = h_{(ab)} + \frac{1}{4} g_{ab} h^{c}_{c},
\eea
where $g^{\mu \nu} H_{(\mu \nu)} = g^{ab} h_{(ab)} = 0$ and we have performed a linearised Weyl shift to decouple the FRW graviton from the internal scalars.  We fix the internal gauge freedom by imposing the de Donder-type gauge conditions
\begin{align}
\nabla^{a} h_{(ab)} &= \nabla^{a} h_{a \mu} = 0.
\end{align}

With these gauge choices, we can expand the metric fluctuations in spherical harmonics as
\begin{align}
&H_{(\mu \nu)} = H^{I}_{(\mu \nu)} Y^{I}, \quad H_{\mu}^{\mu} =  H^{I} Y^{I}, \quad 
h_{(ab)} = \Phi^{I} Y^{I}_{(ab)}, \cr &h^{a}_{a} =  \pi^{I} Y^{I},  \quad
h_{\mu a} =  B^{I}_{\mu} Y^{I}_{a}, 
\end{align}
with the summation over $I$ denoting a sum over all possible spherical harmonics. 

When $\dot{\phi} = 0$, the $S^4$ becomes independent of the FRW space and the background Riemann tensors may be written as
\begin{align}
\label{nc_riemann}
&R_{\mu \nu \rho \sigma} = e^{4 \phi} H_{c}^2 \left( g_{\mu \rho} g_{\nu \sigma} - g_{\nu \rho} g_{\mu \sigma} \right), \quad
R_{abcd} = e^{-2 \phi} L_0^{-2} ( g_{ac} g_{bd} - g_{ad} g_{bc}),
\end{align}
where we have used $\dot{a}(t) = H_{c} a(t)$, with $H_c$ denoting the constant Hubble parameter at the end of inflation.
In general, the linearised Ricci tensor may be written
\bea
\delta R_{MN}(h_{MN}) &=& -\frac{1}{2} \biggl[ \Box_8 h_{MN} + \nabla_{M} \nabla_{N} h^{P}_{P} - \nabla_{M} \nabla^{P} h_{PN} - \nabla_{N} \nabla^{P} h_{P M} \nn
&-&2 R_{MPQN} h^{PQ} - R_{M}^{~P} h_{NP} - R_{N}^{~P} h_{MP} \biggr].
\eea
As we are interested in examining the stability at the minimum when $\dot{\phi} = 0$, we can write the d'Alembertian operator in d=8 in terms of the sum of  d'Alembertian operators on the FRW spacetime and $S^4$, $\Box_8 = \Box_{FRW} + \Box_{S^4}$, where $\Box_{FRW} \equiv g_{FRW}^{\mu \nu} \nabla_{\mu} \nabla_{\nu}$ and $\Box_{S^4} \equiv g_{S^4}^{ab} \nabla_{a} \nabla_{b}$. We will also use $- e^{2 \phi} L_0^2 \Box_{S^{4}}  \equiv \lambda^{I} Y^{I}$, where the eigenvalues $\lambda^{I}$ of the ordinary Laplacian on $S^4$ for the various tensor harmonics are 
\be
\begin{array}{ccc}
 \mbox{Harmonic} & \lambda^{I} & \mbox{Range of}~ $k$\\
Y^{I} & k(k+3) & k \geq 0 \\
Y^{I}_{a} & k(k+3) -1 & k \geq 1 \\
Y^{I}_{(ab)} & k(k+3)-2 & k \geq 2
\end{array}
\ee 
Note as $\Box_{S^4}$ always makes a contribution of the correct sign, we focus on the stability of the lowest mode (smallest $k$) in each case.

We start by considering the $S^4$ variation $\delta R_{ab} = 8 \pi G \delta T_{ab}$. The left hand side is
\bea
-\frac{1}{2}  \biggl[ (\Box_{FRW} + \Box_{S^4}) h_{(ab)} &+& \frac{1}{4} g_{ab} (\Box_{FRW} + \Box_{S^4}) h^{c}_{c} + \nabla_{a} \nabla_{b} H^{\rho}_{\rho} - \nabla_{a} \nabla^{\rho} h_{\rho b} - \nabla_{b} \nabla^{\rho} h_{\rho a} \nn &-& \frac{8 }{L_0^2 e^{2 \phi}} h_{(ab)} - \frac{3}{2} \nabla_{a} \nabla_{b} h^{c}_{c}  \biggr],
\eea
while $8 \pi G \delta T_{ab}$ is
\begin{align}
16 \pi G &\biggl[ \frac{1}{12} \Lambda h_{(ab)} + \frac{1}{48} g_{ab} h^{c}_{c} \left(\Lambda - \frac{3}{q^2 L_0^4 e^{4 \phi}} \right) \biggr].
\end{align}
Decomposing in terms of spherical harmonics, one immediately sees that
\bea
\left[\left(\Box_{FRW} + \Box_{S^4} - \frac{8}{3} \pi G \left( \frac{3}{q^2 L_0^4 e^{4 \phi} } - \Lambda \right) \right)\pi^{I}  \right] Y^{I} &=& 0,  \label{eqn1} \\
 \left( H^{I} - \frac{3}{2} \pi^{I} \right)\nabla_{(a} \nabla_{b)} Y^{I} &=& 0, \label{eqn2} \\
(\nabla^{\mu} B_{\mu}^{I} ) \nabla_{(a} Y^{I}_{b)} &=& 0, \\
\left[ (\Box_{FRW} + \Box_{S^4}) - \frac{8}{ L_0^2 e^{2 \phi}} + \frac{8}{3} \pi G \Lambda \right] \Phi^{I} Y^{I}_{(ab)} &=& 0 \label{eqn4}.
\eea
For certain low-lying scalar harmonics, some or all of their derivatives appearing above may vanish. As our concern here is stability and establishing a window in parameter space, we consider the generic case where all the derivatives of $Y^{I}$ are non-zero. In effect, the coefficients in the above equations must all vanish independently. From (\ref{eqn1})-(\ref{eqn4}) we find that $H^{I}$ and $\pi^{I}$ are not independent, $B_{\mu}^{I} $ is transverse, and that the mass squared $m^2$ associated to $\pi^{I}$ and $\Phi^{I}$ are non-tachyonic provided  
\bea
\Lambda &\leq& \frac{3e^{-4 \phi} }{q^2 L_0^4}, \; \; \mbox{or}\; \;  bc \leq \frac{27}{4} [ 1+\sqrt{1-4bc/27}]^2, \\
\Lambda &\leq& \frac{6 e^{-2 \phi} }{4 \pi G L_0^2}, \; \; \mbox{or} \;\; bc \leq 36 [ 1+\sqrt{1-4bc/27}]. 
\eea
Neither of these constraints reduce the window of allowable parameters (\ref{window}) noted earlier. 

We now turn our attention to the mixed components of the Einstein equation $\delta R_{\mu a} = 8 \pi G \delta T_{\mu a}$. We find these become
\bea
&-&\frac{1}{2} \biggl[(\Box_{FRW} + \Box_{S^4}) h_{\mu a} + \frac{3}{4} \nabla_{\mu} \nabla_{a} H^{\rho}_{\rho} -\frac{3}{4} \nabla_{\mu} \nabla_{a} h^{d}_{d}  - \nabla_{\mu} \nabla^{\nu} h_{\nu a} - \nabla_{a} \nabla^{\nu} H_{(\mu \nu)} \nn &-& 3 \left( e^{4 \phi} H_c^{2} + \frac{e^{-2 \phi}}{L_0^2} \right) h_{\mu a} \biggr] 
= - \frac{4 \pi G}{3} \left( \frac{3 e^{-4 \phi} }{2 q^2 L_0^4} - \Lambda\right) h_{\mu a}.
\eea
This in turn means
\bea
\left[ (\Box_{FRW} + \Box_{S^4})  - \nabla_{\mu} \nabla^{\nu} \delta^{~ \mu}_{\nu}  - 3\left(e^{4 \phi} H_c^2 + \frac{e^{-2 \phi}}{L_0^2} \right) - \frac{8 \pi G}{3}\left( \frac{3 e^{-4 \phi}}{2 q^2 L_0^4} - \Lambda\right)
 \right] B^{I}_{\mu} Y_{a}^{I} &=& 0, \nn
 \left[  -\nabla^{\nu} H^{I}_{(\mu \nu)} \nabla_{a} + \frac{3}{4} \nabla_{\mu} H^{I} \nabla_{a}  - \frac{3}{4} \nabla_{\mu} \pi^{I} \nabla_{a}   \right] Y^{I} &=& 0. \nonumber 
\eea
The upper  equation above, when evaluated at the compactification radius, places looser stability constraint derived from the graviphoton fluctuation $B_{\mu}^{I}$
\bea
\Lambda \leq \frac{3e^{-4 \phi}}{2 q^2 L_0^4} + \frac{9 e^{-2 \phi}}{4 \pi GL_0^2} + \frac{9 e^{4 \phi} H_c^2}{8 \pi G}. 
\eea
Even when $H_c=0$, this condition may be re-written in terms of $b$ and $c$ as \be bc \leq \frac{27}{8} [ 1+ \sqrt{1-4bc/27}] [5 + \sqrt{1-4 bc/27}].\ee This again, does not constrain the parameters considered earlier. 

Finally we consider the FRW part $\delta R_{\mu \nu } = 8 \pi G \delta T_{\mu \nu}$
\begin{align}
-&\frac{1}{2} \biggl[ (\Box_{FRW} + \Box_{S^4}) H_{\mu \nu} + \nabla_{\mu} \nabla_{\nu} H^{\rho}_{\rho} - \nabla_{\mu} \nabla^{\rho} H_{\rho \nu} - \nabla_{\nu} \nabla^{\rho} H_{\rho \mu} - 2 R_{\mu \rho \sigma \nu} H^{\rho \sigma} \nn & - R_{\mu}^{~\rho} H_{\rho \nu} - R_{\nu}^{~\rho} H_{\rho \mu} \biggr] + \frac{1}{4} g_{\mu \nu} (\Box_{x} + \Box_{y}) h^{c}_{c} \nn
&= 16 \pi G \left[ -\frac{1}{12}\left( H_{\mu \nu} - \frac{1}{2} g_{\mu \nu} h^{c}_{c} \right)\left( \frac{3 e^{-4 \phi}}{2 q^{2} L_0^{4}} - \Lambda\right) - \frac{1}{12} g_{\mu \nu} \left(-\frac{3 e^{-4 \phi}}{4 q^{2} L_0^{4}} h^{c}_{c}\right)\right],
\end{align}
or in terms of spherical harmonics that
\begin{align}
\label{ext_grav}
\biggl[ & \delta R_{\mu \nu} (H^{I}_{\rho \sigma}) - \frac{1}{2} \Box_{S^4} H^{I}_{\mu \nu} + \frac{4 \pi G}{3} \left( \frac{3 e^{-4 \phi}}{2 q^2 L_0^4} - \Lambda\right) H_{\mu \nu}^{I} + \frac{1}{4} g_{\mu \nu} (\Box_{FRW} + \Box_{S^4}) \pi^{I} \nn &- \frac{2 \pi G}{3} g_{\mu \nu}  \left( \frac{3 e^{-4 \phi}}{2 q^{2} L_0^{4}}  - \Lambda\right) \pi^{I} - \frac{\pi G e^{-4 \phi}}{q^{2} L_0^{4}}g_{\mu \nu} \pi^{I} \biggr] Y^{I} = 0.
\end{align}

We now find ourselves in a position to establish the existence of the FRW graviton. The graviton comes from the constant $Y^{I}$ mode of the equation (\ref{ext_grav}). Using (\ref{eqn1}), it may be brought to the form
\be
\delta R_{\mu \nu} (H^{I}_{\rho \sigma}) + \frac{4 \pi G}{3} \left( \frac{3 e^{-4 \phi}}{2 q^2 L_0^4} - \Lambda\right) H_{\mu \nu}^{I} = 0.
\ee
One may readily check that it is the correct equation for the linearised graviton in FRW (\ref{nc_riemann}), by using the equations of motion (\ref{eoms}) to recast it as 
\be
\delta R_{\mu \nu} (H^{I}_{\rho \sigma}) - 3 e^{4 \phi} H_c^2 H_{\mu \nu}^{I} = 0.
\ee
One may then simply proceed as in \cite{DeWolfe:2001nz} by defining a shifted massive tensor field $\varphi_{(\mu \nu)}^{I} = H_{(\mu \nu)}^{I} + C \nabla_{(\mu} \nabla_{\nu)} \pi^{I}$, where one may determine the constant $C$ so that $\varphi_{(\mu \nu)}^{I}$ is transverse $\nabla^{\mu} \varphi_{(\mu \nu)}^{I} = 0$. The contribution to the mass from the spherical harmonic eigenvalues have the correct sign, so these modes are also stable. 

\section{Conclusions}
Evoking extra-dimensional Einstein gravity with a positive cosmological constant and a reduction on a compact space with a Yang-Mills instanton, we have obtained an effective potential dependent on two parameters $b,c$ which offers much promise as an inflationary model. One neat feature is that the size of the compact space is dynamically stabilised during the inflationary process. 

We have explored the parameter space for models of inflation consistent with current observational bounds from CMB data, and have identified some attractive features. In particular, as the inflaton starts off rolling from a local maximum, our model falls into the class of hilltop models, and as such, any number of efolds may be accommodated classically. 

We have also looked at metric fluctuations to see how the requirement that the final vacuum at the local minimum be stable, affects the allowed window of the parameters $b, c$. We find that it does not constrain the permissible values further. In general, the fluctuations on the Yang-Mills instanton should also be incorporated into the analysis, but treating spherical harmonics for non-Abelian gauge fields is quite involved. We recognise that the analysis here is not complete and would like to see if stability is still guaranteed in the presence of both gauge and metric fluctuations in future work. 

To fulfill the potential of this set-up as a viable model of inflation, we need to consider the introduction of secondary fields for both slow-roll ($bc \gtrsim 6.5$) and fast-roll inflation ($bc \lesssim 6.5$). In the former case, we require a waterfall field to ensure reheating, while in the latter the inflaton has to be liberated from the role of generating the curvature perturbations, so that observational bounds are not violated. Modifying the higher dimensional action by including higher order corrections, or including a warp-factor in the reduction ansatz, offer means of introducing extra scalars. It would be interesting to address these in future work. 

Another exciting area of immediate future investigation is generalisations to different dimensions. For simplicity, we considered the Yang-Mills instanton in a $(4+4)$-dimensional set-up. In general, to stabilise a topological configuration of a Yang-Mills field in 
dimensions greater than four, we need higher order terms of the gauge field strength. Any meaningful comparison between compactifications from different dimensions would hopefully identify a mechanism for picking out the four-dimensions of our observed universe. 

\section*{{\large Acknowledgements}}
We are grateful to Bin Chen, Oliver DeWolfe, Qing-Guo Huang, Misao Sasaki, Hossein Yavartanoo, Piljin Yi and Patta Yogendran for discussions. In particular, we would also like to thank Prava Chingangbam for participation in earlier discussions, and helpful comments throughout. We are also grateful to Hironobu Kihara for bringing this area to our attention and for comments on the final draft.

\end{document}